\documentclass[letter]{aa}
\usepackage[varg]{txfonts}
\usepackage[english]{babel}
\usepackage{graphicx}
\usepackage{hyperref}
\hypersetup{colorlinks=true, urlcolor=blue, linkcolor=blue, citecolor=blue}
\usepackage{color}

\begin{document}
\title{Unidentified emission features in the R Coronae Borealis star V854~Centauri}
\titlerunning{Unidentified emission features in the RCB star V854~Cen}

\author{L.~C.~Oostrum \inst{\ref{api},\ref{astron}}
\and
B.~B.~Ochsendorf \inst{\ref{jhu}}
\and
L.~Kaper \inst{\ref{api}}
\and
A.~G.~G.~M.~Tielens \inst{\ref{leiden}}
}

\institute{Anton Pannekoek Institute for Astronomy, University of Amsterdam, PO Box 94249, 1090 GE, Amsterdam, The Netherlands\label{api} \\ \email{l.c.oostrum@uva.nl}
\and
ASTRON, The Netherlands Institute for Radio Astronomy, PO box 2, 7990 AA Dwingeloo, The Netherlands\label{astron}
\and
Department of Physics and Astronomy, The Johns Hopkins University, 3400 North Charles Street, Baltimore, MD 21218, USA\label{jhu}
\and
Leiden Observatory, Leiden University, PO Box 9513, 2300 RA Leiden, The Netherlands\label{leiden}
}

\date{Received date / Accepted date}

\abstract{During its 2012 decline the R Coronae Borealis star (RCB) V854~Cen was spectroscopically monitored with
X-shooter on the ESO {\it Very Large Telescope}. The obscured optical and near-infrared spectrum exhibits many narrow and several broad emission features, as previously observed. The envelope is spatially resolved along the slit and allows for a detailed study of the circumstellar material. In this {\it Letter} we report on the properties of a number of unidentified emission features (UFs), including the detection of a new one at $8692\,\mathrm{\AA}$. These UFs have been observed in the Red Rectangle, but their chemical and physical nature is still a mystery. The previously known UFs behave similarly in the Red Rectangle and V854~Cen, but are not detected in six other observed RCBs. Possibly the presence of some hydrogen is required for the formation of their carrier(s). The $\lambda$8692 UF is present in all RCBs. Its carrier is likely of a carbonaceous molecular nature, presumably different from that of the other UFs.}


\maketitle

\section{Introduction}
\label{sec:introduction}
R Coronae Borealis stars (RCBs) are rare, hydrogen deficient supergiants that exhibit strong declines in their brightness \citep{clayton2012}. Only about a hundred RCBs are known in the Galaxy. Their rarity is an indication of a short evolutionary phase and/or points to a peculiar mode of stellar evolution: (i) a merger between a CO and He white dwarf or (ii) a final helium-shell flash leading to the expansion to supergiant size \citep{iben1996,saio2002}. The decline is thought to be due to the formation of clouds of carbon dust along the line of sight, obscuring the stellar photosphere so that the circumstellar envelope becomes observable in emission. Such a natural coronograph provides a unique opportunity to study the chemical and physical nature of the circumstellar envelope of these peculiar objects. 

Due to the irregularity of these events, limited spectra are available of RCBs in decline. V854~Cen is an RCB that is of particular interest because it is one of the few RCBs that include hydrogen lines in their spectra; It is the most hydrogen-rich RCB after DY~Cen \citep{asplund1998,jeffery1993}. V854~Cen and DY~Cen are also the only RCBs in which polycyclic aromatic hydrocarbons (PAHs) have been detected \citep{gh2011a}. The spectra include the weak 18.9$\,\mathrm{\mu m}$ band that is now generally attributed to C$_{60}$ \citep{cami2010,sellgren2010}. Besides that, V854~Cen shows some unidentified visual emission features (UFs) in its decline spectrum \citep{rao1993b}. These features have only been detected in the Red Rectangle proto-planetary nebula \citep{schmidt1991}. In that object, the features change in shape, intensity and peak position as a function of position in the nebula \citep{vanwinckel2002,wehres2011}. Additional impetus for a study of the visual emission features is provided by the potential link between the visual emission bands in the Red Rectangle (RR) and the diffuse interstellar band (DIB) absorption features in the interstellar medium \citep{sarre1995}.

In this {\it Letter} we show, for the first time, the spatial structure of the emission features in V854~Cen during its 2012 decline. We compare the characteristics of these features to those detected in the RR. In addition, we search for new emission features (300--2500$\,$nm) in V854~Cen, as well as in six other RCBs.

\section{Observations and data reduction}
\label{sec:observations}
Time on the ESO VLT was granted for observing V854~Cen within a window of four months in 2012. The object was monitored by the American Association of Variable Star Observers (AAVSO). When the star's visual magnitude dropped below $m_\mathrm{v}=8$ (maximum-light $m_\mathrm{v}=7.1$), multiple spectra were obtained with \mbox{VLT/X-shooter} \citep{vernet2011} during the decline that lasted around 2.5 months. Additionally, in 2013 X-shooter spectra were taken of V854~Cen and a few other RCBs, known to be in decline as determined from their AAVSO light curves, with the aim to search for the presence of unidentified features. V854~Cen was then at maximum light. For all observations, the highest resolution mode was used, where $R\approx10000$, 18000, and 11500 for the UVB, VIS, and NIR arms, respectively. A log of observations is given in Table~\ref{tab:obs_overview}.

The spectra were reduced using the \mbox{X-shooter} pipeline version 2.2.0 \citep{modigliani2010} and flux calibrated using spectrophotometric standard stars. Telluric correction of the 1D NIR spectra was done with Spextool \citep{vacca2003} using telluric standard spectra obtained at similar airmass and close in time to the targets. The resulting spectra were shifted to the rest-frame of the observed source.

\begin{table}
\caption{Log of VLT/X-shooter observations. During the rise of V854~Cen, three different position angles were used. The magnitudes were measured from the acquisition images, except for V854~Cen during maximum light, for which they are saturated. For these, values from the AAVSO were used. V854 Cen distance is based on the $(V-I)$ - $M_V$ relation for RCBs in the LMC by \mbox{\citet{tisserand2009}}. For the others, a typical value of $M_\mathrm{v}=-5$ \citep{clayton2012} is assumed. The signal-to-noise ratio was measured in the continuum around 6200~\AA.}
\label{tab:obs_overview}
\centering
\small
\begin{tabular}{p{1.4cm} l p{.5cm} l p{.9cm} l l }
\hline\hline\\[-8pt]
Target & $d$ & Phase & Obs. date & $m_\mathrm{v}$ & $m_\mathrm{v}$ & $S/N$ \\
 & (kpc) & & & (max) & (obs) & \\\hline\\[-9pt]
V854~Cen & 2.4  & Max & 2012-05-04 & 7.1$^{(1)}$ & 7.2 & 90 \\ 
 & & Min & 2012-06-14 & & 13.3 & 70 \\ 
 &  & Min & 2012-06-18 & & 12.8 & 130 \\ 
 &  & Rise & 2012-07-05 & & 9.9 & 170 \\  
 &  & Rise & 2012-07-05 & & 9.9 & 150 \\  
 &  & Rise & 2012-07-05 & & 9.9 & 170 \\  
 &  & Max & 2013-07-15 & &  7.2 & 85 \\
NSV 8092  & 17.5 & Min & 2013-07-15 & $<$11.7$^{(2)}$ & 14.0 & 30 \\
R~CrB & 1.3 & Min & 2013-07-15 & 5.7$^{(3)}$ & 13.7 & 35 \\
RT~Nor & 9.9 & Min & 2013-07-15 & 10.6$^{(3)}$ & 15.2 & 20 \\
RZ~Nor & 10.5 & Min & 2013-07-15 & 10.6$^{(3)}$ & 16.2 & 15 \\
S~Aps & 7.1 & Min & 2013-07-15 & 9.6$^{(3)}$ & 14.3 & 10 \\
V~CrA & 6.6 & Min & 2013-07-15 & 9.4$^{(3)}$ & 17.5 & 10 \\
\hline
\end{tabular}
\tablebib{(1) \citet{samus2003}; (2) \citet{tisserand2013}; (3) \citet{ducati2002}.}
\end{table}

\section{The unidentified emission features}
\label{sec:band_properties}
\subsection{Detected bands}

\begin{figure*}
\centering
\includegraphics[width=18cm]{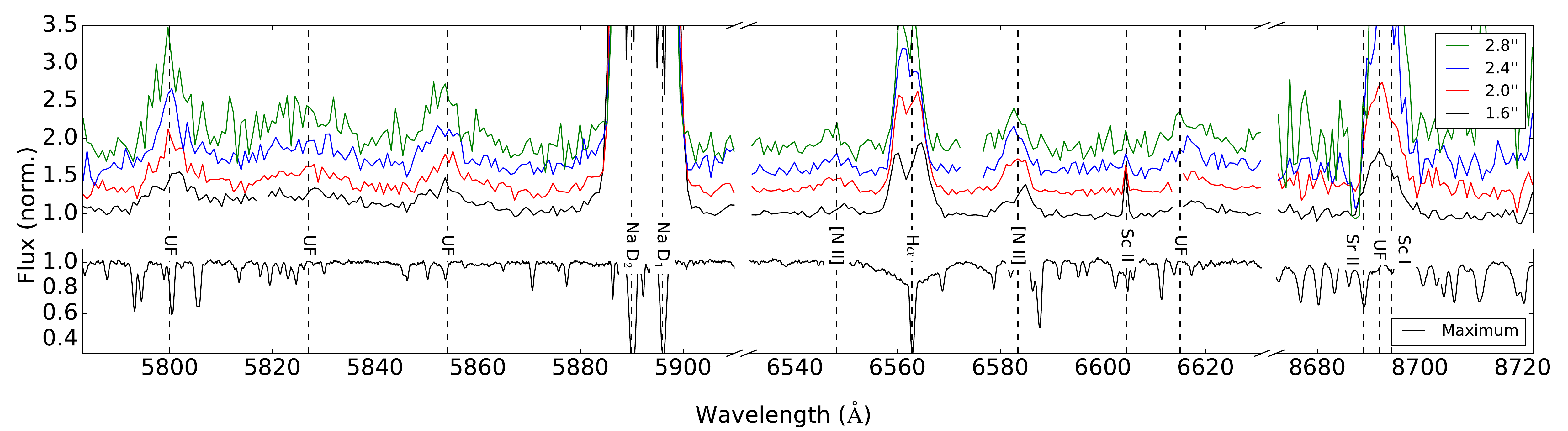}
\caption{V854~Cen decline spectra at different offsets from the central object (\emph{see inset}) and at maximum light (\emph{bottom spectrum}). During a decline, the absorption spectrum converts into an emission spectrum, the strongest emission lines being the Na~{\sc i}~D doublet. Additionally, a sequence of unidentified emission features (UFs) is clearly detected between 5800 and 5860$\mathrm{\,\AA}$, at 6617$\mathrm{\,\AA}$, and at 8692$\mathrm{\,\AA}$. 
The narrow Y~{\sc ii} $\lambda6614$ emission line has been removed from the $1.6\arcsec$ and $2.0\arcsec$ offset spectra for plotting purposes. The absorption lines near $5800\,\mathrm{\AA}$ and $5854\,\mathrm{\AA}$ are due to C~{\sc i} and Ba~{\sc ii}, respectively.}
\label{fig:v854_spectrum}
\end{figure*}

A section of the \mbox{X-shooter} spectrum of V854~Cen is shown in Fig.~\ref{fig:v854_spectrum}. The broad features at 5800, 5827, 5854 and 6617~\AA\ were first recognized in V854~Cen by \citet{rao1993b} during a deep decline ($m_V\sim15$). The similarity between these features and the emission features in the RR was already noted. We expand upon this by considering the spatial and dynamical (i.e. radial velocity) structure of these features in V854~Cen. None of these bands are detected in any of the other observed RCBs. The spectra reveal the presence of a broad emission feature at 8692~\AA\ that has not been seen before in any RCB, nor in the RR.

Of the seven bands, only the $\lambda\lambda$5800, 5827, 5854, and 6617 bands are detected in the 2012 decline of V854~Cen. The absence of three of the features (at 5772, 6774 and 6997$\mathrm{\,\AA}$) may be attributed to the depth of the decline, as this decline is 2 magnitudes -- or a factor ${\sim} 6$ in flux -- shallower than the 1992 decline, and the non-detected features are the weakest ones. The four detected bands are present in the two spectra taken during the deepest part of the decline ($m_V\sim13$), just beyond the stellar continuum along the slit. The features are strongest in the spectrum taken just after the minimum, on 2012-06-14. Only in that spectrum the S/N in the features is high enough to allow for an analysis of their spatial distribution. The features are detected at different distances from the central object as shown in Fig.~\ref{fig:v854_spectrum}, which also shows the maximum light spectrum for comparison. The features are only very marginally detected at the Western side of the star (negative offsets), hence we focus on the Eastern side (positive offsets). None of the bands are detected in the other RCBs.

\begin{figure}
\centering
\includegraphics[width=\columnwidth]{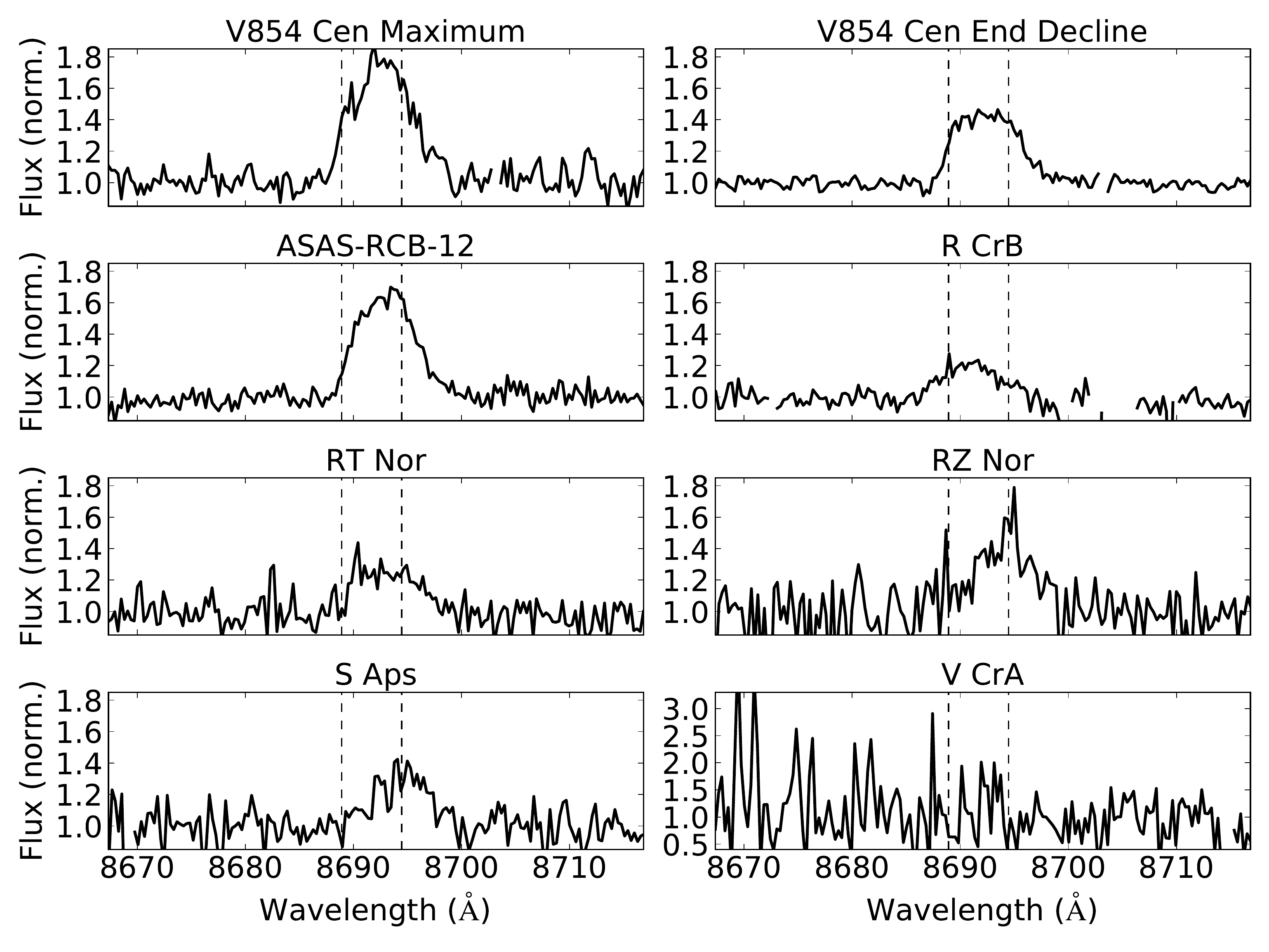}
\caption{The $\lambda8692$ UF in all observed RCBs. The two dashed lines indicate the position of the Sr~{\sc ii} $\lambda8689$ and Sc~{\sc i} $\lambda8694$ photospheric lines. The UF is detected in all except V CrA. The position and intensity of the feature clearly vary between objects. In V854~Cen the feature is also detected during maximum light.}
\label{fig:8692_all}
\end{figure}

The whole wavelength range of \mbox{X-shooter} (300-2500$\,\mathrm{nm}$) has been searched for the presence of emission features. One new feature was detected, present in \emph{all} V854~Cen spectra, at ${\sim}8692\,\mathrm{\AA}$. It is also detected in five of the six other observed RCBs in decline (Fig.~\ref{fig:8692_all}). Similar to many features associated with the circumstellar material, this feature is only detected off-source. The feature is blended with two photospheric lines: Sr~{\sc ii} $\lambda8689$ and Sc~{\sc i} $\lambda8694$. In order to disentangle the nebular features from the stellar ones, the off-source spectrum is divided by the on-source spectrum and the resulting spectrum renormalised. The presence of the feature during maximum-light is especially noteworthy.

\subsection{Spatial and dynamical structure of the bands}
\begin{figure}
\centering
\vbox{
\includegraphics[width=\columnwidth]{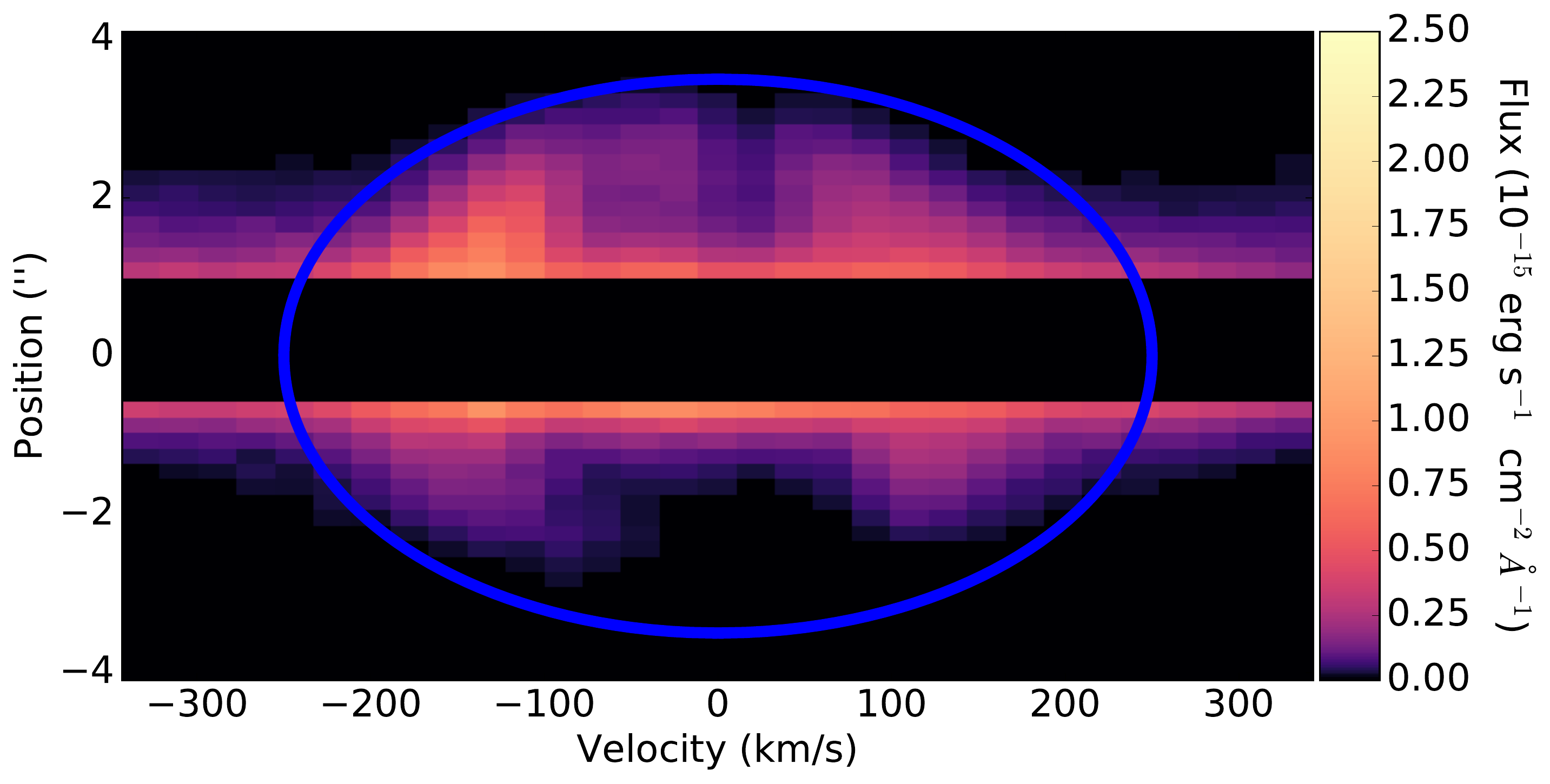}
\includegraphics[width=\columnwidth]{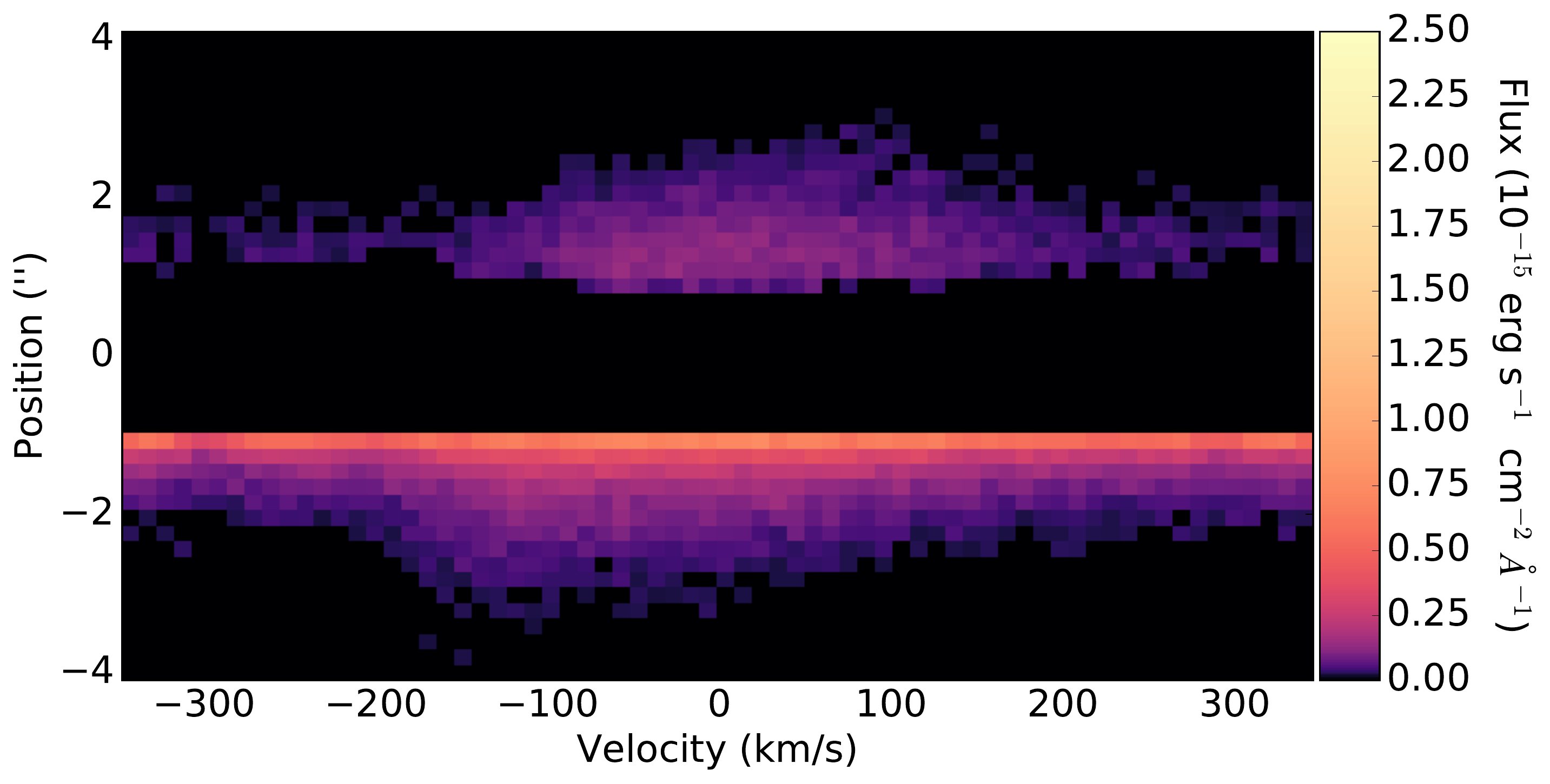}
}
\caption{Position along the spectrograph slit (vertical direction) against radial velocity for the  Ca~{\sc II}~K resonance line ({\it top}) and UF at 8692~\AA\ ({\it bottom}) in V854~Cen.
The continuum flux has been subtracted and all flux within 0.7\arcsec\ of the central object has been set to zero to enhance the visibility of these extended features. 
The thick blue line shows the expected maximum radial velocity at each position for a spherical shell of radius 3.5\arcsec\ and expanding with a velocity of $250\mathrm{\,km\,s^{-1}}$. One sees that the unidentified feature at 8692$\mathrm{\,\AA}$ is spatially extended as well.}
\label{fig:2D}
\end{figure}

As the circumstellar envelope of V854 is spatially resolved along the spectrograph slit, position-velocity (PV) diagrams can be produced to study the kinematic structure of the emission lines as a function of the distance from the star. In Fig.~\ref{fig:2D} we show the dynamical structure as revealed by the Na~{\sc i} D resonance lines, and UF8692. The other unidentified features are too weak to construct a PV diagram. Both doublet components in Na~{\sc i}~D show a structure suggesting that the emission is produced in roughly a shell with a radius of about 3.5\arcsec\ and expanding with a velocity of $250\mathrm{\,km\,s^{-1}}$. The size of the shell is based on the furthest position where the flux is more than 3$\sigma$ above the noise. PV diagrams of more extended features, confirming the size and velocity of the shell, are shown in Appendix~\ref{app:pv}. The shape of the UF8692 extended emission is different from the shell traced by Na~{\sc i}~D. It shows a change in wavelength and width for different distances from the central object. For V854 Cen, it shifts toward the red at the Eastern side of the star (i.e. positive offsets), and to the blue at the Western side of the star. Additionally, it narrows with increasing distance to the star on the Eastern side. It is also worth noting that the equivalent width (W$_\mathrm{eq}$) of the feature -- here defined such that an emission feature has a positive W$_\mathrm{eq}$ -- increases with distance to the star. This confirms that this newly found feature originates from the circumstellar material.

The behaviour of the $\lambda8692$ feature is different for each RCB. In RT~Nor and RZ~Nor, the width of the feature does not change significantly, while a narrowing with increasing distance is observed in the others. The position of the feature shifts in all objects. There is no clear pattern to this, in some objects only blueshift is observed, in others also redshift and/or no shift on one side of the star. As the RCBs have different distances, different physical scales are probed, but we do not find a correlation between the behaviour of this feature and the respective distances to the RCBs.

The shape of the bands, when integrated over an interval along the slit, is well described by either one or two Gaussians. All features in the $\lambda5825$ complex are measured together, as those bands are too close in wavelength to be considered separately. For the band at $5800\,\mathrm{\AA}$, two Gaussians are used to account for the asymmetric band shape, which significantly improves the fits. For the other bands a single component is used. All sharp emission lines, originating from regions close to the star \citep[${\sim}2R_*$;][]{clayton1996} instead of the large-scale circumstellar material, are removed from the spectra prior to the fitting procedure.

For all five detected bands, the measured peak position as function of distance to the star is shown in Fig.~\ref{fig:band_shift} (black squares). All features (at 1.6$\arcsec$ and 2.8$\arcsec$) are shown in Appendix~\ref{app:uf}. The narrow component in the $5800\,\mathrm{\AA}$ feature (Fig.~\ref{fig:v854_spectrum}) is used for the band position, as this component is stronger and dominates the peak position. The same bands in the RR are known to show a blueshift with increasing distance to the central binary system \citep{vanwinckel2002,wehres2011}. The same analysis method as for V854~Cen is applied to the RR spectra, except for the number of components: The higher S/N RR spectra require two components for the $\lambda5854$ and $\lambda6617$ bands in addition to $\lambda5800$, whereas one component is sufficient for those bands in V854~Cen. The only RR spectrum covering the $\lambda8692$ feature is from the ESPaDOnS spectrograph mounted on the CFHT, and unfortunately has insufficient signal in that wavelength region (N.L.J.~Cox, priv. comm.). The measured RR band positions agree well with those provided by \citet{wehres2011}. With the exception of $\lambda8692$, all V854~Cen and RR bands show a shift towards shorter wavelengths with increasing distance to the central object. The RR bands are clearly shifted towards the blue with respect to V854~Cen, although they are known to be redder than the V854~Cen bands close to the central binary in the RR \citep{vanwinckel2002}. The $\lambda8692$ feature is the only feature detected on both sides of the star, and shows a redshift where the other features show a blueshift.

The RR bands show a correlation between their position and width: Some bands become narrower as they shift to shorter wavelengths \citep{schmidt1991,sarre1995,vanwinckel2002,wehres2011}. We are unable to confirm such a correlation in the V854~Cen features. It should be noted that, in the RR, the features change most rapidly close to the central binary. The distance to V854~Cen is roughly three times that of the RR \citep[$710\mathrm{\,pc}$;][]{menshchikov2002}. We thus cannot exclude the presence of a width-wavelength correlation in the inner regions of the circumstellar material.

\begin{figure}
\centering
\includegraphics[width=9cm]{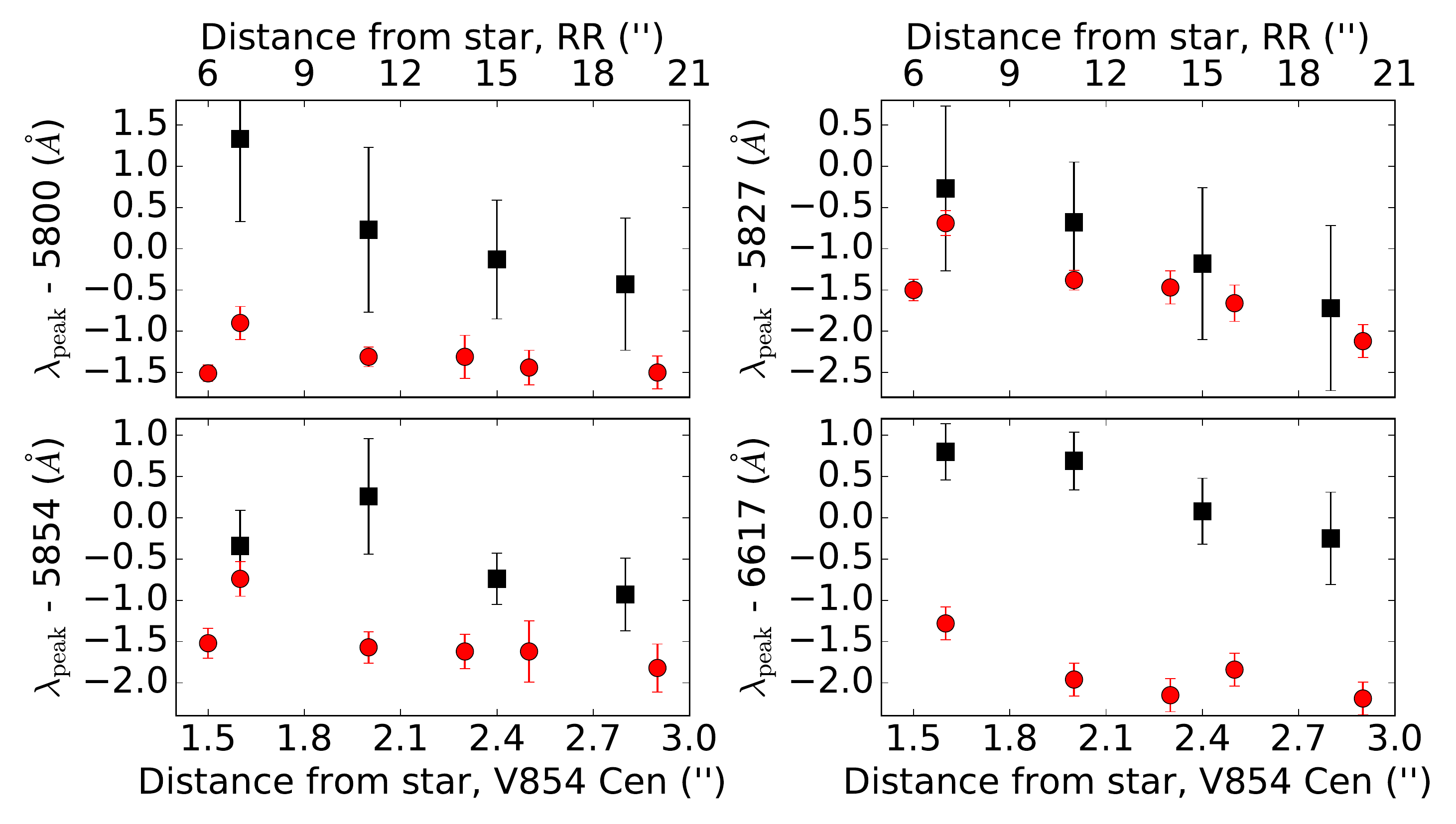}
\includegraphics[width=9cm]{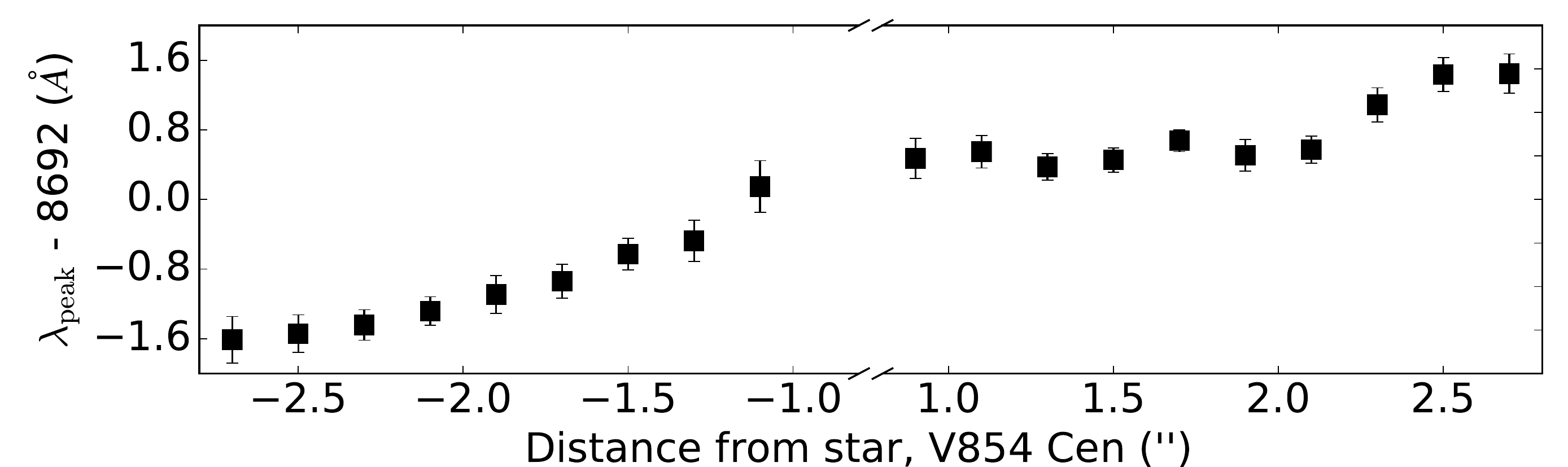}
\caption{Band positions of the unidentified features as function of distance to the central object for V854~Cen (black squares) and the Red~Rectangle (red circles). There is no measurement of the $\lambda6617$ feature at $6\arcsec$. Error bars reflect 1$\sigma$ errors. All RR bands are blueshifted with respect to the corresponding bands in V854~Cen. The $\lambda8692$ band is only covered in V854~Cen and is detected on both sides of the star.}
\label{fig:band_shift}
\end{figure}

\section{Discussion}
\label{sec:discussion}
We detect several emission features in the nebula surrounding V854~Cen, up to $3.5\,\arcsec$ or ${\sim}8500\,\mathrm{AU}$ from the central object. This is the furthest detection of the nebula so far. The broad emission lines from Na~{\sc i}~D, H$\alpha$, and [C~{\sc i}] $\lambda9850$ show a roughly spherical outflow with velocities up to $250\,\mathrm{km\,s^{-1}}$. In addition to these atomic lines, we detect six unidentified emission features, including a new one at $8692\,\mathrm{\AA}$. The width of the emission features is comparable to that of other broad emission lines. The UFs that are also detected in the RR have a similar width, even though the velocities in the RR are much lower \citep[${\sim}7\,\mathrm{km\,s^{-1}}$ in CO;][]{jura1995} than in V854~Cen. This indicates that the features are intrinsically broad and hence that their width in V854~Cen is not due to Doppler broadening. The observed shift of ${\sim}1\,\mathrm{\AA}$ in the features at $5800-5854\,\mathrm{\AA}$ corresponds to ${\sim}50\,\mathrm{km\,s^{-1}}$. This shift may be due to rotational cooling, as in the RR \citep{wehres2011}, however assuming no cooling it is an upper limit on the radial velocity of the carrier. This velociy is significantly lower than that of the Na~{\sc i}~D-traced shell, hence the carriers are not located in that shell. There is, however, evidence for low-velocity dust in RCBs, assuming that the minimum-light optical spectrum is dominated by scattered light from the central object \citep{gh2011b}. This low-velocity material is the most likely environment for the carrier of the UFs. The absence of the features in other RCBs may indicate that the presence of some hydrogen is required for their formation.

The $\lambda$8692 feature may provide insight in the geometry of this dust in V854~Cen, as it is the only UF that is detected clearly on both sides of the star. If we interpret the shift in this feature as a Doppler shift, it is consistent with a bipolar outflow which is being accelerated to ${\sim}50\,\mathrm{km\,s^{-1}}$. A bipolar geometry has been suggested in literature \citep[e.g.][]{rao1993a,chesneau2014}. If the other UFs are part of the same outflow, we would expect them to show a shift of ${\sim}0.3\,\mathrm{\AA}$ between their closest and furthest detection. We are, however, unable to confirm this, as this putative shift is smaller than the detected shifts in both V854~Cen and the RR. By the same argument as for the other features, the carrier of $\lambda8692$ is not located in the high-velocity shell.

For the observed bands, the shifts in the RR are consistent with a change in excitation temperature of a molecule \citep{wehres2011}. For V854~Cen too, one would expect such a shift as the temperature in the outflow decreases.

Care has to be taken when comparing UF8692 to the other features, as it is the only feature present in all spectra and the only one to show up in other RCBs. It is thus unlikely that this feature has the same carrier as one of the other emission bands. The presence of the feature in more hydrogen-deficient RCBs does indicate that the carrier is hydrogen-poor. A carbonaceous molecular nature seems likely, given the high carbon abundances of RCBs. The different behaviour of the feature in different RCBs is quite puzzling. It may be an indication of varying physical and chemical conditions in the complex circumstellar envelopes of RCBs.

It has been suggested that the UFs in the RR are the emission equivalent of the DIBs \citep[e.g.][]{scarrott1992}. The inverted DIB spectrum does show correlations with some UFs. We are unable to identify a possible DIB counterpart for the $\lambda8692$ feature. The closest known DIB is located at $8648.3$~\AA\ (online catalogue\footnote{\url{http://leonid.arc.nasa.gov/DIBcatalog.html}} and N.L.J.~Cox, priv. comm.), which is too far off to be the likely absorption equivalent of the $\lambda8692$ feature.

\begin{acknowledgements}
The authors thank Nadine Wehres and Hans van Winckel for providing us the Red Rectangle spectra. We acknowledge the variable star observations from the AAVSO International Database contributed by observers worldwide, and used in this research. LCO would like to thank Nick Cox for discussion about DIBs and the Red Rectangle. LCO acknowledges funding from the European Research Council under the European Union's Seventh Framework Programme \mbox{(FP/2007-2013)} / ERC Grant Agreement n. 617199. Based on observations collected at the European Organisation for Astronomical Research in the Southern Hemisphere under ESO programmes \mbox{089.D-0937(A)} and \mbox{091.C-0934(B)}.
\end{acknowledgements}

\bibliographystyle{aa}
\bibliography{v854}

\begin{thebibliography}{27}
\expandafter\ifx\csname natexlab\endcsname\relax\def\natexlab#1{#1}\fi

\bibitem[{{Asplund} {et~al.}(1998){Asplund}, {Gustafsson}, {Rao}, \&
  {Lambert}}]{asplund1998}
{Asplund}, M., {Gustafsson}, B., {Rao}, N.~K., \& {Lambert}, D.~L. 1998, \aap,
  332, 651

\bibitem[{{Cami} {et~al.}(2010){Cami}, {Bernard-Salas}, {Peeters}, \&
  {Malek}}]{cami2010}
{Cami}, J., {Bernard-Salas}, J., {Peeters}, E., \& {Malek}, S.~E. 2010,
  Science, 329, 1180

\bibitem[{{Chesneau} {et~al.}(2014){Chesneau}, {Millour}, {De Marco}, {Bright},
  {Spang}, {Lagadec}, {M{\'e}karnia}, \& {de Wit}}]{chesneau2014}
{Chesneau}, O., {Millour}, F., {De Marco}, O., {et~al.} 2014, \aap, 569, L4

\bibitem[{{Clayton}(1996)}]{clayton1996}
{Clayton}, G.~C. 1996, \pasp, 108, 225

\bibitem[{{Clayton}(2012)}]{clayton2012}
{Clayton}, G.~C. 2012, Journal of the American Association of Variable Star
  Observers (JAAVSO), 40, 539

\bibitem[{{Ducati}(2002)}]{ducati2002}
{Ducati}, J.~R. 2002, VizieR Online Data Catalog, 2237

\bibitem[{{Garc{\'{\i}}a-Hern{\'a}ndez}
  {et~al.}(2011{\natexlab{a}}){Garc{\'{\i}}a-Hern{\'a}ndez}, {Kameswara Rao},
  \& {Lambert}}]{gh2011a}
{Garc{\'{\i}}a-Hern{\'a}ndez}, D.~A., {Kameswara Rao}, N., \& {Lambert}, D.~L.
  2011{\natexlab{a}}, \apj, 729, 126

\bibitem[{{Garc{\'{\i}}a-Hern{\'a}ndez}
  {et~al.}(2011{\natexlab{b}}){Garc{\'{\i}}a-Hern{\'a}ndez}, {Rao}, \&
  {Lambert}}]{gh2011b}
{Garc{\'{\i}}a-Hern{\'a}ndez}, D.~A., {Rao}, N.~K., \& {Lambert}, D.~L.
  2011{\natexlab{b}}, \apj, 739, 37

\bibitem[{{Iben} {et~al.}(1996){Iben}, {Tutukov}, \& {Yungelson}}]{iben1996}
{Iben}, Jr., I., {Tutukov}, A.~V., \& {Yungelson}, L.~R. 1996, \apj, 456, 750

\bibitem[{{Jeffery} \& {Heber}(1993)}]{jeffery1993}
{Jeffery}, C.~S. \& {Heber}, U. 1993, \aap, 270, 167

\bibitem[{{Jura} {et~al.}(1995){Jura}, {Balm}, \& {Kahane}}]{jura1995}
{Jura}, M., {Balm}, S.~P., \& {Kahane}, C. 1995, \apj, 453, 721

\bibitem[{{Men'shchikov} {et~al.}(2002){Men'shchikov}, {Schertl}, {Tuthill},
  {Weigelt}, \& {Yungelson}}]{menshchikov2002}
{Men'shchikov}, A.~B., {Schertl}, D., {Tuthill}, P.~G., {Weigelt}, G., \&
  {Yungelson}, L.~R. 2002, \aap, 393, 867

\bibitem[{{Modigliani} {et~al.}(2010){Modigliani}, {Goldoni}, {Royer},
  {Haigron}, {Guglielmi}, {Fran{\c c}ois}, {Horrobin}, {Bristow}, {Vernet},
  {Moehler}, {Kerber}, {Ballester}, {Mason}, \& {Christensen}}]{modigliani2010}
{Modigliani}, A., {Goldoni}, P., {Royer}, F., {et~al.} 2010, in \procspie, Vol.
  7737, Observatory Operations: Strategies, Processes, and Systems III, 773728

\bibitem[{{Rao} \& {Lambert}(1993{\natexlab{a}})}]{rao1993a}
{Rao}, N.~K. \& {Lambert}, D.~L. 1993{\natexlab{a}}, \aj, 105, 1915

\bibitem[{{Rao} \& {Lambert}(1993{\natexlab{b}})}]{rao1993b}
{Rao}, N.~K. \& {Lambert}, D.~L. 1993{\natexlab{b}}, \mnras, 263, L27

\bibitem[{{Saio} \& {Jeffery}(2002)}]{saio2002}
{Saio}, H. \& {Jeffery}, C.~S. 2002, \mnras, 333, 121

\bibitem[{{Samus'} {et~al.}(2003){Samus'}, {Goranskii}, {Durlevich}, {Zharova},
  {Kazarovets}, {Kireeva}, {Pastukhova}, {Williams}, \& {Hazen}}]{samus2003}
{Samus'}, N.~N., {Goranskii}, V.~P., {Durlevich}, O.~V., {et~al.} 2003,
  Astronomy Letters, 29, 468

\bibitem[{{Sarre} {et~al.}(1995){Sarre}, {Miles}, \& {Scarrott}}]{sarre1995}
{Sarre}, P.~J., {Miles}, J.~R., \& {Scarrott}, S.~M. 1995, Science, 269, 674

\bibitem[{{Scarrott} {et~al.}(1992){Scarrott}, {Watkin}, {Miles}, \&
  {Sarre}}]{scarrott1992}
{Scarrott}, S.~M., {Watkin}, S., {Miles}, J.~R., \& {Sarre}, P.~J. 1992,
  \mnras, 255, 11P

\bibitem[{{Schmidt} \& {Witt}(1991)}]{schmidt1991}
{Schmidt}, G.~D. \& {Witt}, A.~N. 1991, \apj, 383, 698

\bibitem[{{Sellgren} {et~al.}(2010){Sellgren}, {Werner}, {Ingalls}, {Smith},
  {Carleton}, \& {Joblin}}]{sellgren2010}
{Sellgren}, K., {Werner}, M.~W., {Ingalls}, J.~G., {et~al.} 2010, \apjl, 722,
  L54

\bibitem[{{Tisserand} {et~al.}(2013){Tisserand}, {Clayton}, {Welch}, {Pilecki},
  {Wyrzykowski}, \& {Kilkenny}}]{tisserand2013}
{Tisserand}, P., {Clayton}, G.~C., {Welch}, D.~L., {et~al.} 2013, \aap, 551,
  A77

\bibitem[{{Tisserand} {et~al.}(2009){Tisserand}, {Wood}, {Marquette}, {Afonso},
  {Albert}, {Andersen}, {Ansari}, {Aubourg}, {Bareyre}, {Beaulieu}, {Charlot},
  {Coutures}, {Ferlet}, {Fouqu{\'e}}, {Glicenstein}, {Goldman}, {Gould},
  {Gros}, {de Kat}, {Lesquoy}, {Loup}, {Magneville}, {Maurice}, {Maury},
  {Milsztajn}, {Moniez}, {Palanque-Delabrouille}, {Perdereau}, {Rich},
  {Schwemling}, {Spiro}, \& {Vidal-Madjar}}]{tisserand2009}
{Tisserand}, P., {Wood}, P.~R., {Marquette}, J.~B., {et~al.} 2009, \aap, 501,
  985

\bibitem[{{Vacca} {et~al.}(2003){Vacca}, {Cushing}, \& {Rayner}}]{vacca2003}
{Vacca}, W.~D., {Cushing}, M.~C., \& {Rayner}, J.~T. 2003, \pasp, 115, 389

\bibitem[{{Van Winckel} {et~al.}(2002){Van Winckel}, {Cohen}, \&
  {Gull}}]{vanwinckel2002}
{Van Winckel}, H., {Cohen}, M., \& {Gull}, T.~R. 2002, \aap, 390, 147

\bibitem[{{Vernet} {et~al.}(2011){Vernet}, {Dekker}, {D'Odorico}, {Kaper},
  {Kjaergaard}, {Hammer}, {Randich}, {Zerbi}, {Groot}, {Hjorth}, {Guinouard},
  {Navarro}, {Adolfse}, {Albers}, {Amans}, {Andersen}, {Andersen}, {Binetruy},
  {Bristow}, {Castillo}, {Chemla}, {Christensen}, {Conconi}, {Conzelmann},
  {Dam}, {de Caprio}, {de Ugarte Postigo}, {Delabre}, {di Marcantonio},
  {Downing}, {Elswijk}, {Finger}, {Fischer}, {Flores}, {Fran{\c c}ois},
  {Goldoni}, {Guglielmi}, {Haigron}, {Hanenburg}, {Hendriks}, {Horrobin},
  {Horville}, {Jessen}, {Kerber}, {Kern}, {Kiekebusch}, {Kleszcz}, {Klougart},
  {Kragt}, {Larsen}, {Lizon}, {Lucuix}, {Mainieri}, {Manuputy}, {Martayan},
  {Mason}, {Mazzoleni}, {Michaelsen}, {Modigliani}, {Moehler}, {M{\o}ller},
  {Norup S{\o}rensen}, {N{\o}rregaard}, {P{\'e}roux}, {Patat}, {Pena}, {Pragt},
  {Reinero}, {Rigal}, {Riva}, {Roelfsema}, {Royer}, {Sacco}, {Santin},
  {Schoenmaker}, {Spano}, {Sweers}, {Ter Horst}, {Tintori}, {Tromp}, {van
  Dael}, {van der Vliet}, {Venema}, {Vidali}, {Vinther}, {Vola}, {Winters},
  {Wistisen}, {Wulterkens}, \& {Zacchei}}]{vernet2011}
{Vernet}, J., {Dekker}, H., {D'Odorico}, S., {et~al.} 2011, \aap, 536, A105

\bibitem[{{Wehres} {et~al.}(2011){Wehres}, {Linnartz}, {van Winckel}, \&
  {Tielens}}]{wehres2011}
{Wehres}, N., {Linnartz}, H., {van Winckel}, H., \& {Tielens}, A.~G.~G.~M.
  2011, \aap, 533, A28

\end{thebibliography}

\newpage
\begin{appendix}
\onecolumn
\section{Position-velocity diagrams}
\label{app:pv}

\begin{figure}[h]
\centering
\begin{tabular}{cc}
\includegraphics[width=8.5cm]{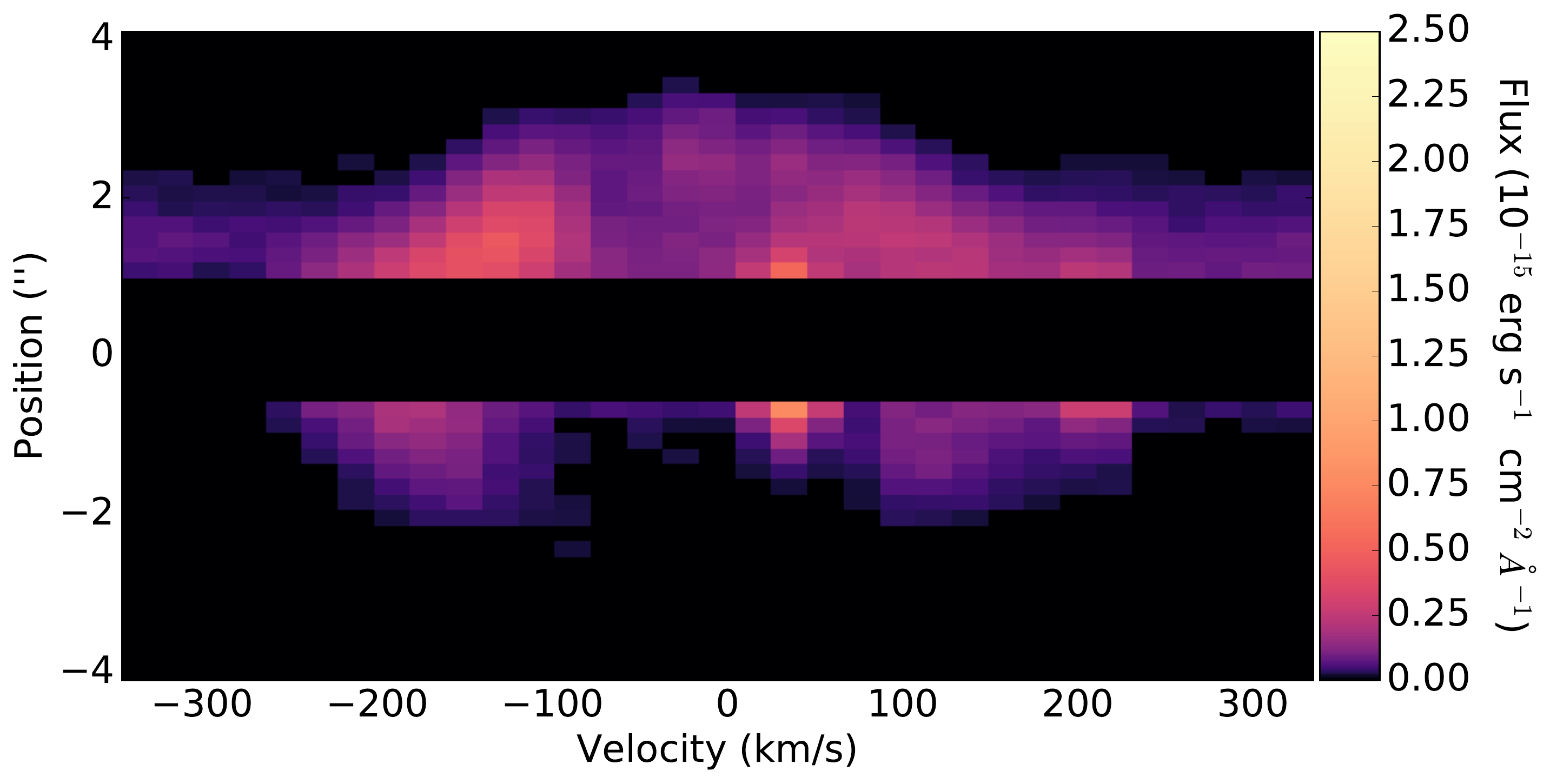} & \includegraphics[width=8.5cm]{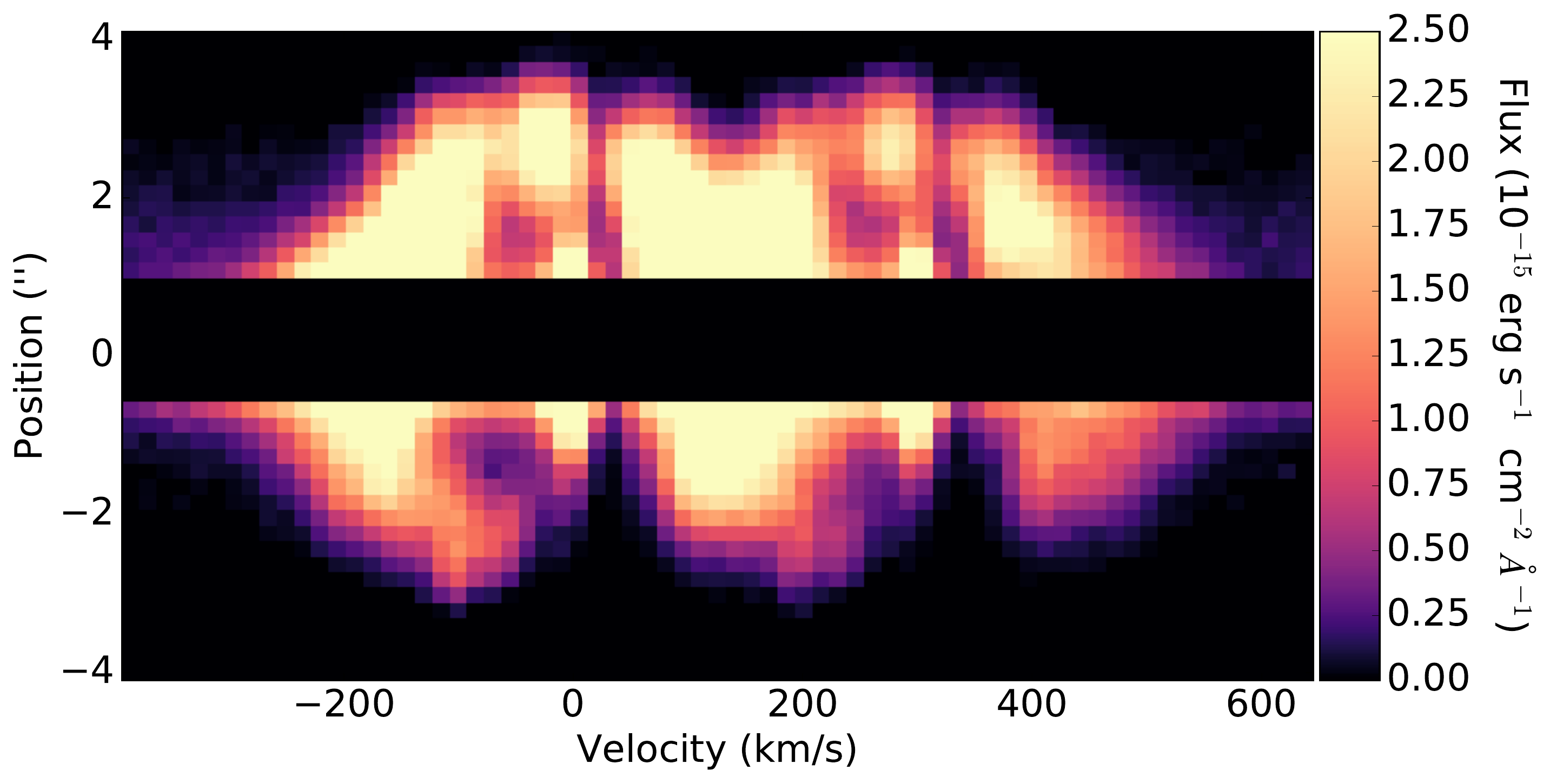} \\
\includegraphics[width=8.5cm]{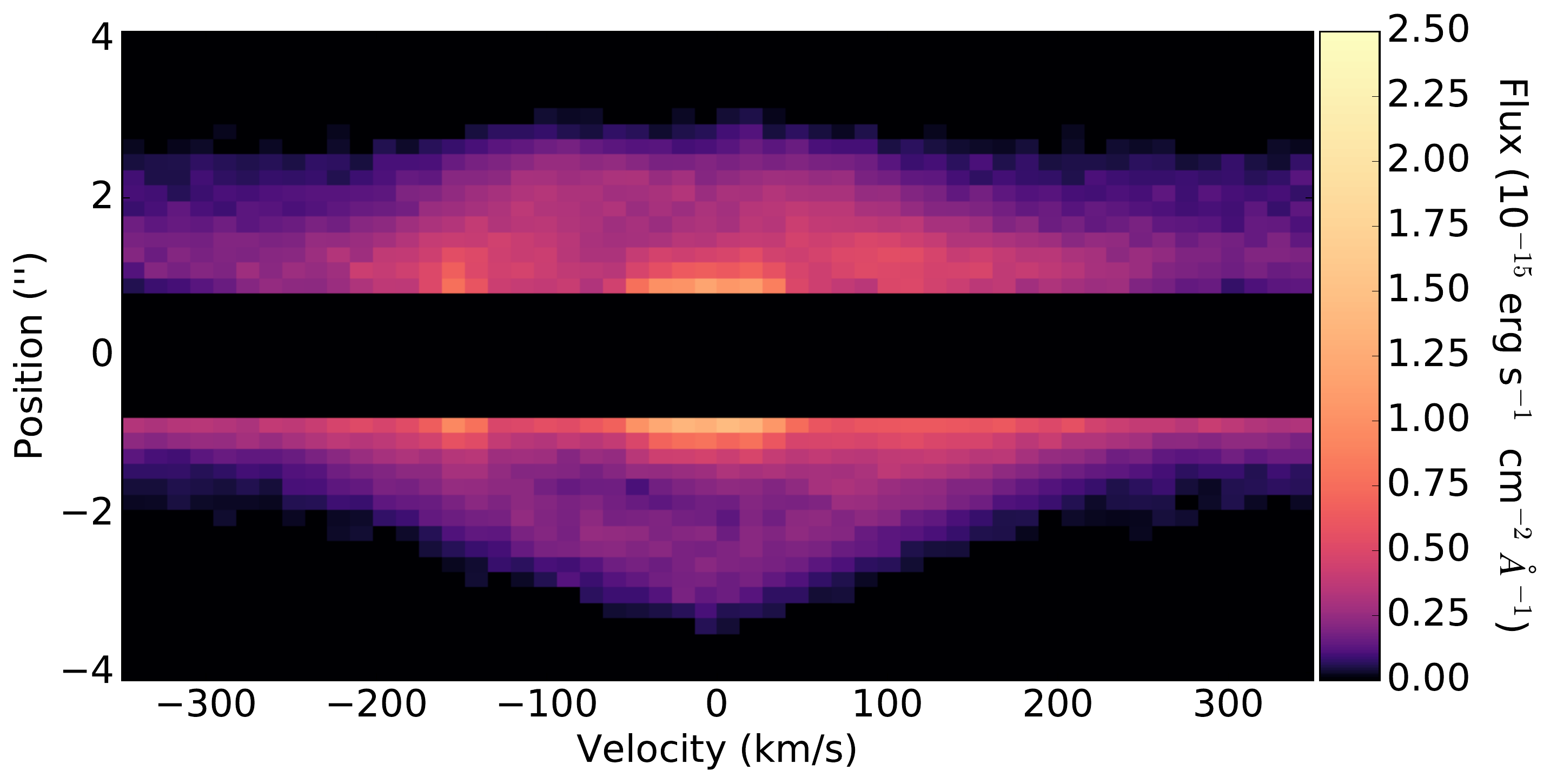} & \includegraphics[width=8.5cm]{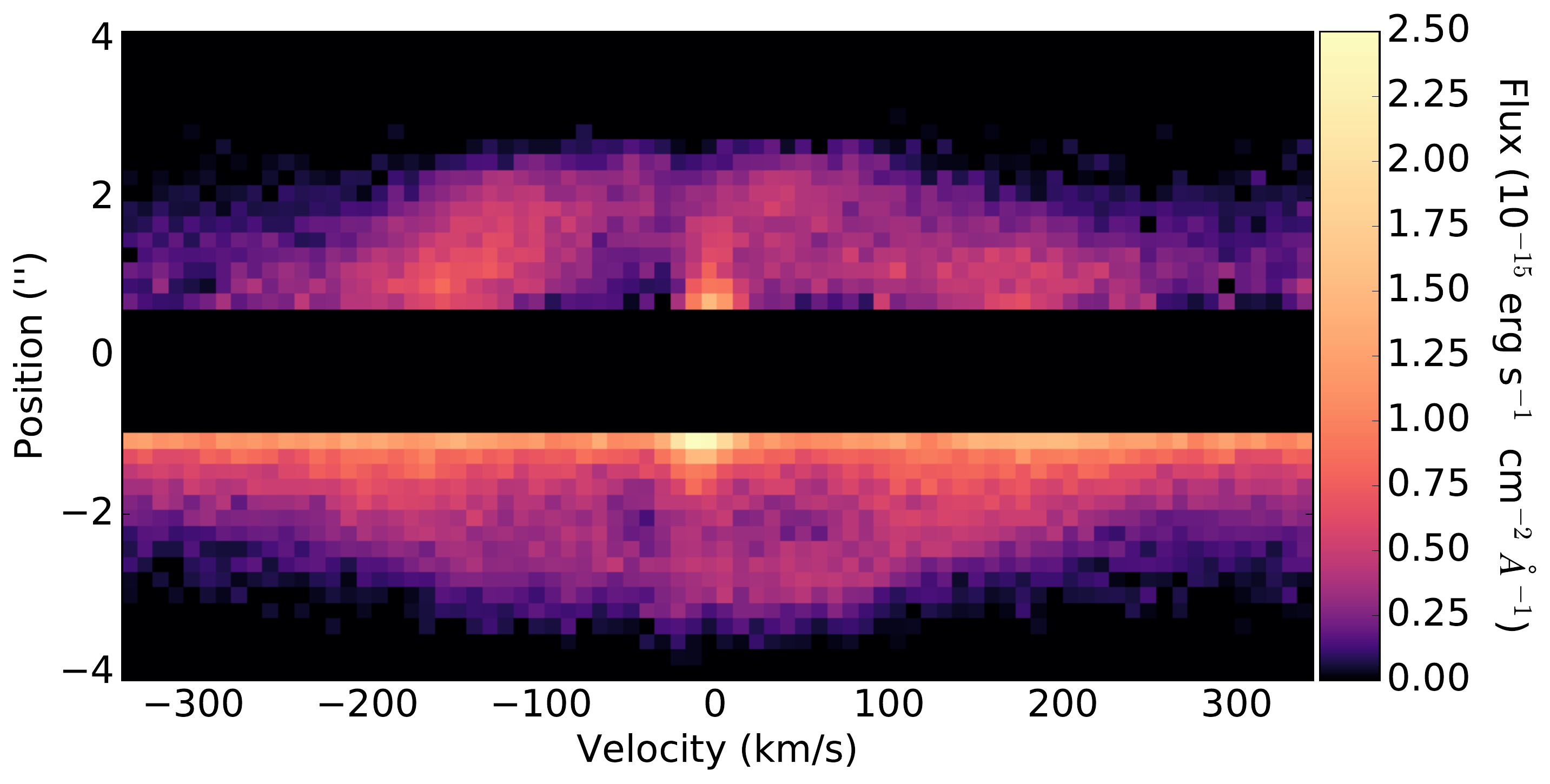} \\
\end{tabular}
\caption{Position-velocity diagrams of CH$^+$ ({\it top left}), Na~{\sc i}~D ({\it top right}), H$\alpha$ ({\it bottom left}) and the [C~{\sc i}] line at 9850~\AA\ ({\it bottom right}) in V854~Cen. Both CH$^+$ and Na~{\sc i}~D show a shape very similar to Ca~{\sc II}~K (see main text); However, both  Na~{\sc i}~D components have a strong zero-velocity component. Note that the narrow emission line in the CH$^+$ diagram at ${\sim}50\,\mathrm{km\,s^{-1}}$ is likely due to Na~{\sc ii}. H$\alpha$ and [C~{\sc i}] show similar, though less pronounced behaviour.}
\end{figure}

\section{Unidentified feature spectra}
\label{app:uf}

\begin{figure}[h]
\centering
\begin{tabular}{cc}
\includegraphics[width=8cm]{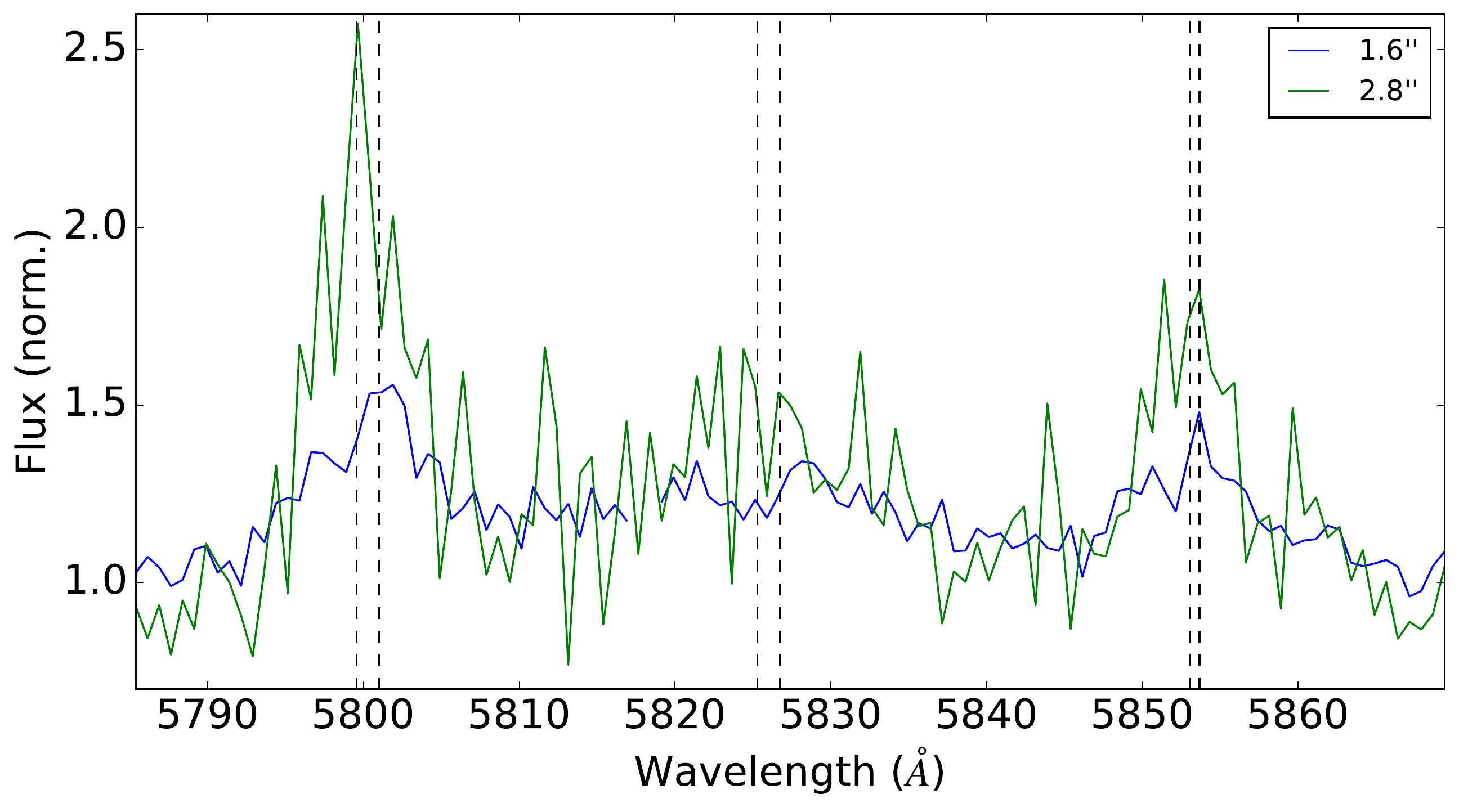} & \includegraphics[width=8cm]{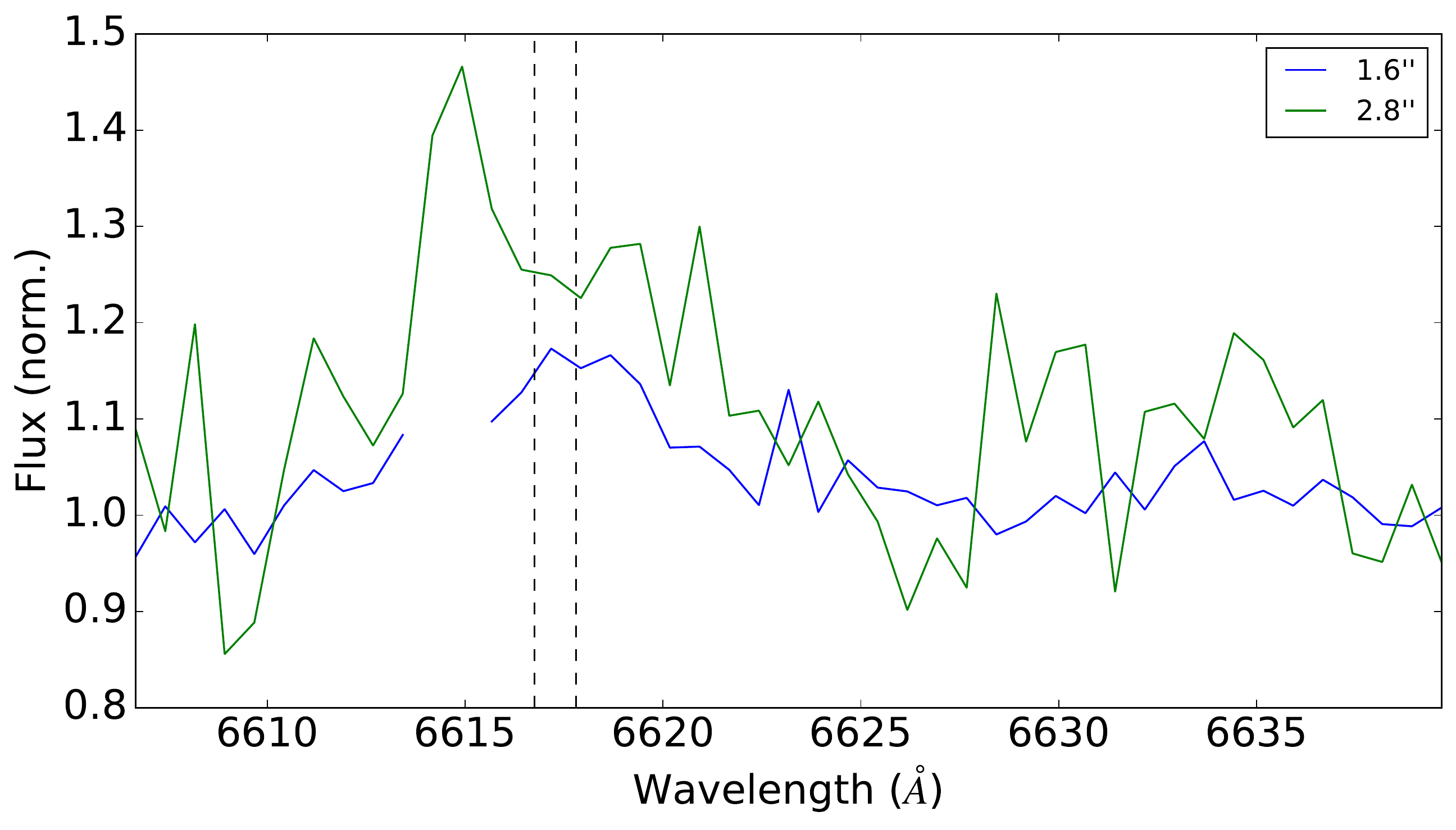} \\
\includegraphics[width=8cm]{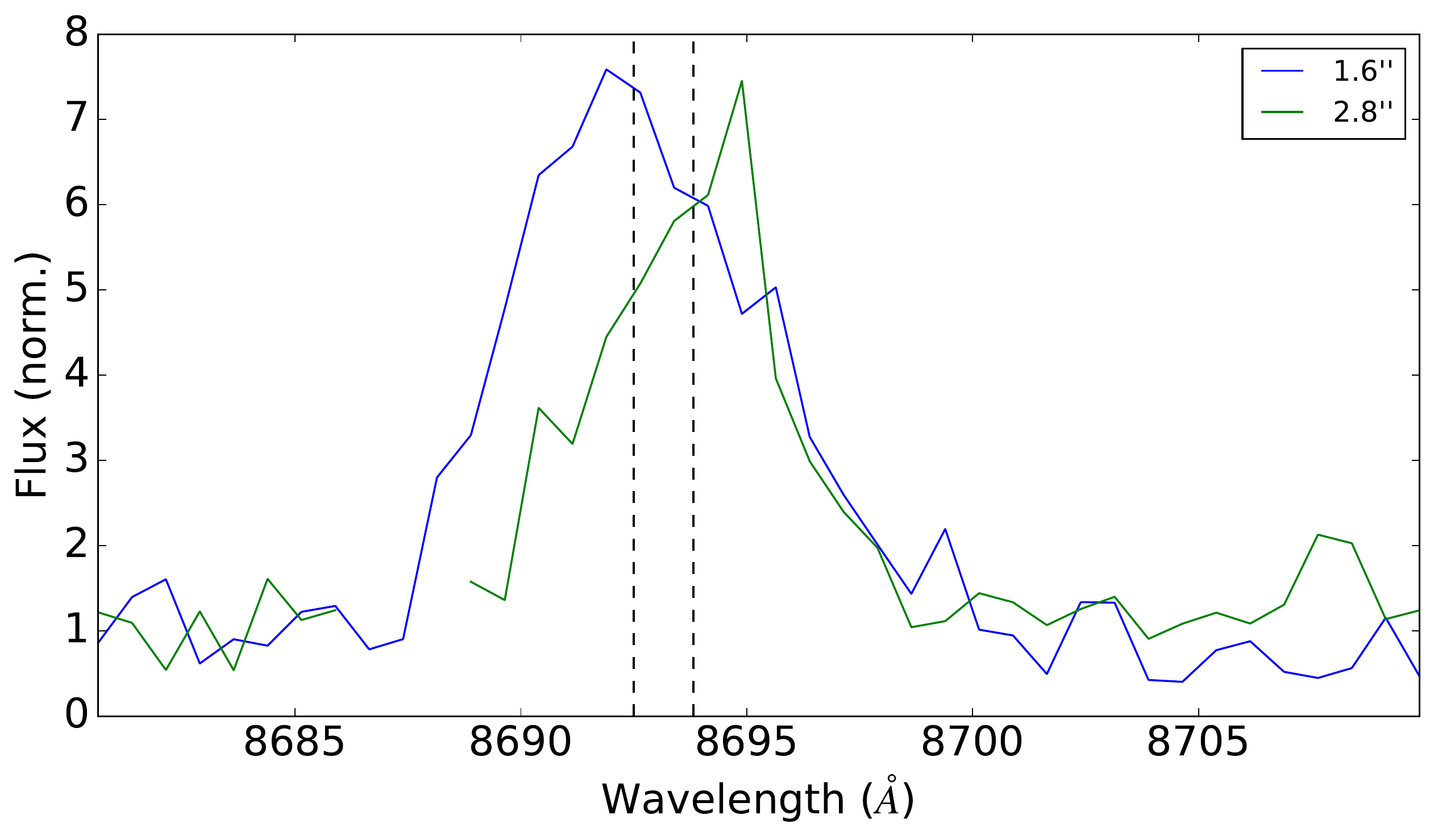} &  \\
\end{tabular}
\caption{Spectra at $1.6\mathrm{\arcsec}$ and $2.4\mathrm{\arcsec}$ from the central object for all unidentified features. Dashed lines indicate the best-fit peak position for each feature. {\it Top left}: $\lambda\lambda$5800, 5827, and 5854. {\it Top right:} $\lambda$6617. {\it Bottom left:} $\lambda8692$. In all features except $\lambda8692$, a blueshift with increasing distance from the central object is observed. The $\lambda$8692 feature shifts toward longer wavelengths in the spectra shown here, but it shifts toward shorter wavelengths on the other side of the star. The $1.6\,\mathrm{\arcsec}$ spectrum of $\lambda$8692 has been stretched by a factor 8 for displaying purposes.}
\end{figure}
\end{appendix}

\end{document}